\documentstyle[epsf,12pt]{article}
\newtheorem{proposition}{Proposition}
\newtheorem{definition}{Definition}
\title{Description of  moduli space of projective 
structures via fat graphs}
\author{V.V.Fock\thanks{on leave of absence from Institute of 
Theoretical and Experimental Physics,\newline 117259, 
B.Cheremushkinskaya 25, Moscow, Russia. }\\
{\em  Institute of Theoretical Physics}\\
{\em Box 803, S-751 08}\\
{\em Uppsala, Sweden}\\E-mail: fock@grotte.teorfys.uu.se}
\date{
\setlength{\unitlength}{\baselineskip}
\begin{picture}(0,0)(0,0)
\put(9,13){\makebox(0,0){UUITP-2/93}}
\put(8,12){\makebox(0,0){hepth/9312193}}
\end{picture}
3 January 1993}
\begin{document}
\newcommand{\sh}{\mbox{sh}}
\maketitle
\begin{abstract}
We give an elementary explicit construction of cell 
decomposition of the moduli space of projective 
structures on a two dimensional surface, analogous 
to the decomposition of Penner/Strebel for moduli space of complex 
structures. The relations between projective structures 
and $PGL(2,{\bf C})$ flat connections are also described.
\end{abstract}
\newpage
The moduli space of projective structures is in a sense 
a phase space of conformal field theories. After the well 
known paper of Kontsevich \cite{K}  the role of 
the Penner/Strebel construction for cell decomposition of 
moduli spaces of complex structures \cite{P,St} was realized 
in understanding relations between the old approach to string 
theory {\em via} conformal field theories and the matrix 
models of nonperturbative gravity (see also \cite{Ch}). 
Briefly the Penner/Strebel 
construction is as follows. 
Let $\Sigma$ be a two dimensional real surface of genus $g$ with $n$
punctures,   ${\cal M}_{g,n}$ be the moduli space of complex structures on 
$\Sigma$. The Penner /Strebel construction gives an isomorphism for 
$g+n \geq 3$,  $n > 0$:
\begin{equation}
 {\cal M}_{g,n} \times {\bf R_+}^n \rightarrow   {\cal M}_{g,n}^{comb},
\end{equation}
where ${\cal M}_{g,n}^{comb}$ is the space of fat graphs, 
homotopicaly equivalent to $\Sigma_0$ with real positive 
numbers assigned to each edge.
( Henceforth we denote by $\Sigma_0$ the surface $\Sigma$ 
with removed punctures.) This construction provides us with 
cell decomposition of 
${\cal M}_{g,n} \times {\bf R_+}^n$ with real 
coordinates for each cell. 

Here we shall consider a slightly different moduli space: 
the moduli space of projective structures. This space is 
in a sense the {\em phase} space for conformal field theories, 
inasmuch as the moduli space of complex structures is the 
{\em configuration} space for these theories.
It turns out, that for our case the construction for cell 
decomposition gives complex coordinates on the moduli 
space of projective structures and can be described by 
means of rather elementary mathematical tools. We also 
consider the relations between the moduli spaces of 
projective structures and of flat 
$PGL(2,{\bf C})$-connections, which are known to be 
very similar. Our approach allows to show that really 
the moduli space of projective structures is indeed 
a blown up covering of the moduli space of flat 
$PGL(2,{\bf C})$-connections. As a by-product we 
get a parameterization of Fuchsian groups. 

I am very indebted to A.A.Rosly for many stimulating 
discussions and to prof. A. Degasperis
for his kind hospitality at the University of Rome, 
where this paper was written.

\section{Generalities on projective structures.}
Let us describe for completeness the notion of 
projective structure on Riemann
surface. The definition of projective structure 
is analogous to that of  
a complex structure . We have only  to replace 
holomorphic functions to M\"obius ones as follows.

A complex structure on a surface is defined 
if there is a full set of 
coordinate patches with complex coordinates 
$z_\alpha$ and holomorphic 
transition functions between them:
\begin{equation}
z_\alpha = \phi_{\alpha \beta}(z_\beta), 
~\overline{\partial}\phi_{\alpha \beta} = 0
\end{equation}

A projective structure on a surface is defined if there is a full set of 
coordinate patches with complex coordinates $z_\alpha$ and M\"obius 
transition functions between them:
\begin{equation} \label{Mob}
z_\alpha = \frac{a_{\alpha \beta}z_\beta + b_{\alpha \beta}}
                               {c_{\alpha \beta}z_\beta + d_{\alpha \beta}}
\end{equation}

A function on a surface equipped with a projective structure 
(a {\em projective surface}) is called {\em projective }
if it can be represented as a M\"obius function of  projective coordinates.
A mapping between two projective surfaces is called {\em projective} if  
it sends projective functions into projective ones, or equivalently if this 
mapping is given by M\"obius functions in projective coordinates.

Projective structure on a surface defines a complex structure on it.  
Conversely, for each complex structure there exists at least one projective 
structure which defines it. Indeed, let  $D$ 
be a unit disk in the complex plane and $\Sigma_0$ be the surface $\Sigma$ 
with removed punctures; and let $p:D \rightarrow \Sigma _0$,  be the 
canonical projection of the universal covering of $\Sigma_0$, 
which by the Poincar\'e uniformization theorem can be identified with the 
unit disk $D$.  Let us take a set of pull-downs of standard projective 
coordinates on $D$ as a set of coordinates on $\Sigma_0$.  The transition 
functions between such coordinates are given by M\"obius functions  thus 
defining a projective structure on $\Sigma_0$. We shall call it the{\em 
Poincar\'e projective structure}. In other words the Poincar\'e projective 
structure is the unique one such that the mapping $p$ is projective w.r.t. it.
The Poincar\'e projective structure is unambiguously defined by the
complex structure of the surface.

 Another example of projective structure can be given in an analogous way 
by Shottky uniformization mapping. We call it the {\em Shottky
projective structure}. This projective structure is unambiguously defined
by the complex structure and the Shottky uniformization data: 
choice of maximal 
set of nonintersecting loops on $\Sigma$. 

The third example of the projective structure on a surface can be given
by representing the surface as a ramified covering of the Riemann sphere.
This projective structure (which we shall call {\em covering 
projective structure}) is well defined outside the ramification points.

The notion of projective structure can be also defined in local terms:
\begin{definition}
Projective connection $T$ on a surface $\Sigma_0$ is a holomorphic section of
a one dimensional complex bundle on $\Sigma_0$ defined by the transition 
functions:
\begin{equation}
T_\alpha(z_\alpha) = T_\beta(z_\beta)\left(\frac{dz_\beta}{dz_\alpha}\right)^2 
+ \frac{1}{2}S(z_\beta, z_\alpha),
\end{equation}
where $S(z_\alpha, z_\beta)$ is the Shwarzian derivative
\begin{equation}
S(z_\alpha, z_\beta) = \left(\frac{d^3z_\alpha}{dz_\beta^3}\right)/
                        \left(\frac{dz_\alpha}{dz_\beta}\right) -
\frac{3}{2}\left(\frac{d^2z_\alpha}{dz_\beta^2}\right)^2/
           \left(\frac{dz_\alpha}{dz_\beta}\right)^2
\end{equation}
\end{definition}

\begin{proposition}
The set of projective structures on a surface compatible with a given 
complex structure is in a bijective correspondence with projective 
connections on the surface.
\end{proposition}

{\em Proof}. Let $T$ be a holomorphic projective connection on the surface
$\Sigma_0$. Consider a ratio of two linearly independent holomorphic  
solutions of the differential equation
\begin{equation}
\partial^2f + Tf = 0. \label{Hil}
\end{equation}
One can easily check that such ratios can be taken as a set of projective
coordinates i.e. that one such ratio is a M\"obius function of another one.
Conversely, let $u_\alpha$ be a full set of projective coordinates on a
surface. Then the expression
\begin{equation}
T_\alpha(z) = S(u_\alpha, z)
\end{equation}
correctly defines a holomorphic projective connection on the surface.
$\Box$

{\em Corrolary}. The moduli space of projective structures 
is an affine bundle over the moduli space of complex structures
${\cal M}_{g,n}$. The fiber of this bundle over a given complex structure
is an affine space over the space of holomorphic quadratic differentials.

Indeed, the difference of two projective connections is a quadratic 
differential and a sum of a projective connection and a quadratic
differential is a projective connection. The existence of at least one
projective connection for each complex structure proves the corrolary.

Note, that the Poincar\'e projective connection gives us a section of this
bundle over ${\cal M}_{g,n}$ and thus we are able to consider this bundle
as a vector bundle. However this section is not holomorphic.
The Shottky projective structure gives us a holomorphic, but multivalued 
section. 

The space of projective structures compatible with a given complex structure 
is an infinite dimensional space provided thai the surface $\Sigma$ has at 
least 
one puncture, since a projective connection $T$ can have arbitrary 
singularities at the punctures. Call a projective structure {\em regular} 
at a puncture $p$ if the corresponding projective connection has at $p$
a pole of order two or less:
\begin{equation}
T(z) = \frac{a}{z^2} + \frac{b}{z} + \mbox{reg. terms},
\end{equation}
where $z$ is a coordinate at a neighborhood of $p$ ($z(p)=0$).

Note that a  projective structure regular at $p$ corresponds to a 
regular at $p$ differential equation (\ref{Hil}) regular at $p$ in the sense 
of Fuchs theory \cite{F}.

\section{Fat graphs and projective surfaces. }
Let us denote by ${\cal MP}_{g,n}$ the space of regular projective 
structures on a surface of genus $g$ with $n$ punctures, (we call
a projective structure regular if it is regular at each puncture 
of the surface); and let ${\cal MP}_{g,n}^{comb}$ be the space of threevalent
fat graphs with positive imaginary part complex numbers assigned to its edges.
\begin{proposition}
An open dense subset of ${\cal MP}_{g,n}$ is isomorphic to
${\cal MP}_{g,n}^{comb}$. 
\end{proposition}
(A rigoristically inclined reader can eliminate the word {\em dense}  from 
the formulation, and consider the present formulation as a conjecture.)

In order to  prove the proposition we shall explicitly construct the mappings 
 ${\cal MP}_{g,n} \rightarrow {\cal MP}_{g,n}^{comb}$ and
${\cal MP}_{g,n}^{comb} \rightarrow {\cal MP}_{g,n}$ and then show
that the former mapping is inverse to the latter.

\subsection{Fat graphs from projective structures}
\label{b}
Let us first describe the mapping ${\cal MP}_{g,n} \rightarrow
{\cal MP}_{g,n}^{comb}$. Let  a {\em projective disk} on $\Sigma$ be
 a mapping 
$u:D \rightarrow \Sigma_0$ of the open standard unit disk equipped with the 
standard projective structure into the surface $\Sigma_0$ considered up
to the action of the group $PGL(2,R)$ of authomorphisms of $D$. Let
${\cal D}_{\Sigma}$ be the set of all projective disks on $\Sigma$.
Define a partial order on ${\cal D}_{\Sigma}$ by taking $u_1 \geq u_2$
($u_1, u_2 \in {\cal D}_{\Sigma}$) if there exists a commutative diagram of 
projective mappings:
\begin{equation}
\setlength{\unitlength}{0.4em}
\begin{picture}(20,20)(0,10)
\put(2,14){\makebox(0,0){$D$}}
\put(2,26){\makebox(0,0){$D$}}
\put(16,20){\makebox(0,0){$\Sigma$}}
\put(5,25){\vector(2,-1){8}}
\put(5,15){\vector(2,1){8}}
\put(2,23){\vector(0,-1){7}}
\put(-1,19){$f$}
\put(8,24){$u_1$}
\put(8,15){$u_2$}
\end{picture}
\end{equation}
Now consider the set ${\cal D}_{\Sigma}^{max} \subset {\cal D}_{\Sigma}$
of maximal disks w.r.t. this ordering. We shall say that a disk 
$u \in {\cal D}_{\Sigma}$ {\em leans} on a puncture $p$ if $p$ belongs
to the closure of image of $u$ in $\Sigma$. In this situation the puncture
$p$ define a discrete set of points $\overline{u}^{-1}(p)$ on the unit circle
$\partial D$. (Here $\overline{u}$ is the extension of $u$ on the closed unit
disk.) Note that it is not necessary that this set consists of one element. 
We shall say that a disc $u$ leans on $p$ with multiplicity 
$\sharp\overline{u}^{-1}(p)$.
In these terms an evident proposition describes the set 
${\cal D}_{\Sigma}^{max}$ of maximal disks:
\begin{proposition}
The set ${\cal D}_{\Sigma}^{max}$ is topologically isomorphic to a graph
with canonical fat graph structure. The vertices of the graph correspond
to projective disks leaning on the punctures at least three times.
\end{proposition}
One can easily see that the set of disks leaning on two given punctures
(or leaning on one given puncture twice) is a one dimensional manifold (fig 1).
\begin{equation}
{\epsfxsize16\baselineskip\epsfbox{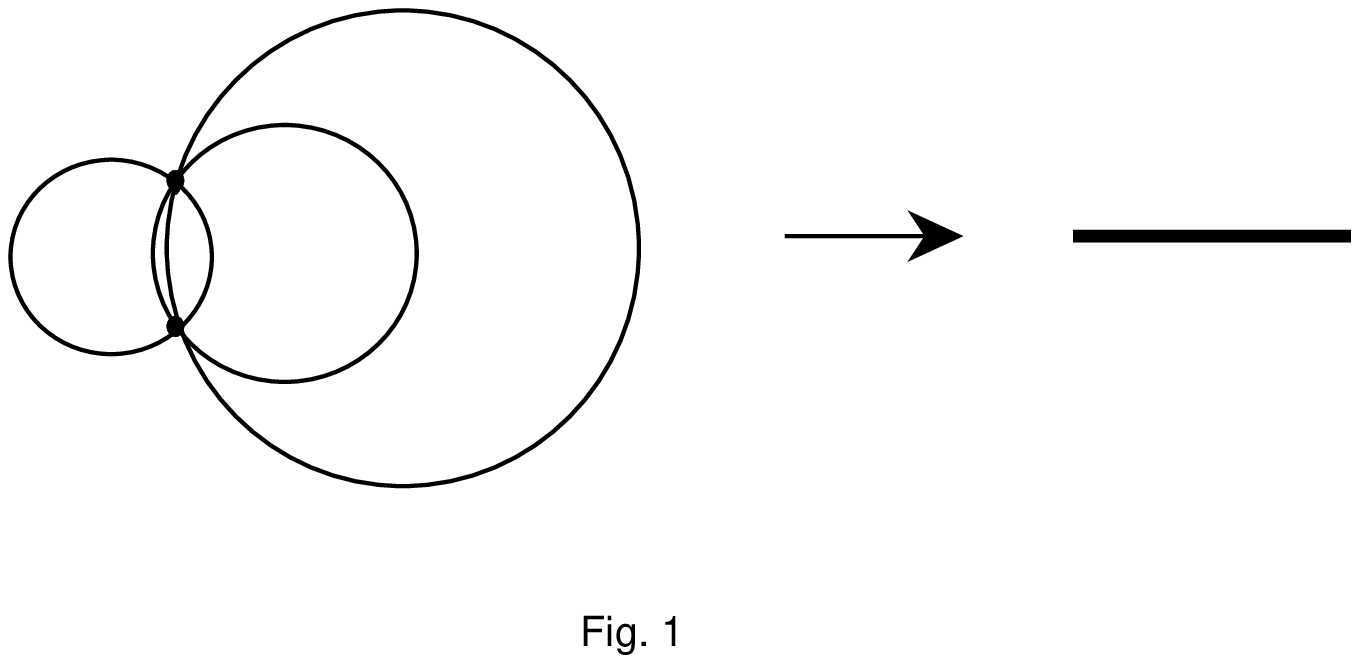}}
\end{equation}
The set of disks close to a disk leaning on $n$ punctures is isomorphic 
to a neighborhood of an $n$-valent vertex of a graph (fig 2).
\begin{equation}
{\epsfxsize16\baselineskip\epsfbox{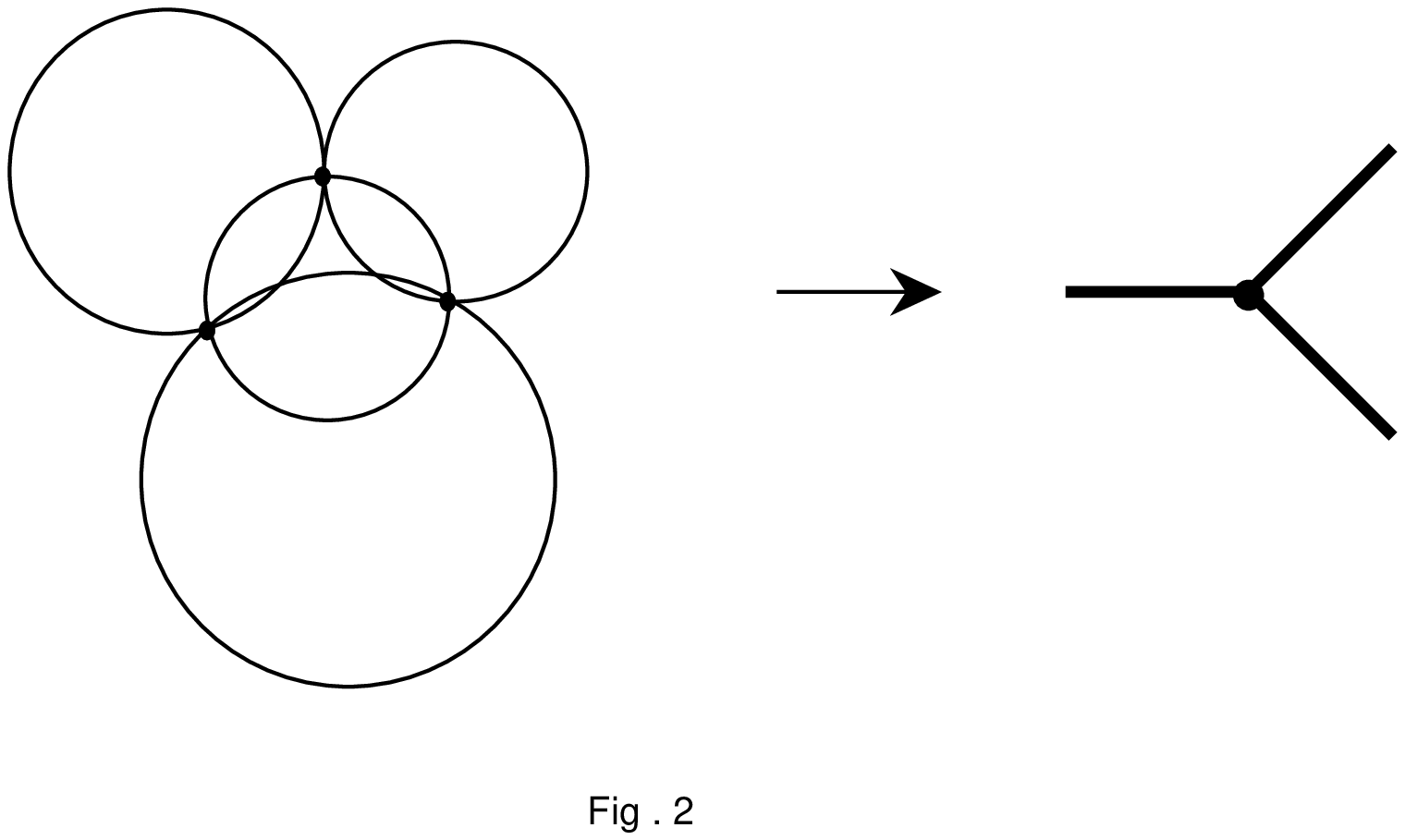}}
\end{equation}
The cyclic
order of the set $\cup_p\overline{u}^{-1}(p)$ on the unit circle induces the
cyclic order of ends of edges incident to the corresponding vertex of 
this graph. $\Box$

Now let us provide ${\cal D}_{\Sigma}^{max}$ with an additional structure --
complex numbers on edges -- which, as it will be demonstrated in the next 
section, contains all the information about the
isomorphism class of projective structure on $\Sigma$.

Let $u_1$ and $u_2$ be two projective disks which correspond to the
beginning and to the end of an oriented edge $\alpha$ respectively.
Let $z_\alpha$ be a (multivalued) projective function on $\Sigma$ such that 
(i) it is equal to
$-1$, $0$, $\infty$ at the points the disk $u_1$ leans on,  
(ii) its imaginary part is negative within the disk,
(iii) The disk $u_2$ leans on points $z_\alpha=0$ and $z_\alpha=\infty$.
These conditions define the function $z_\alpha$ unambiguously. 
(In order to avoid  multivaluedness one can consider here the disks, 
projective functions, e.t.c. on the universal covering $\tilde{\Sigma_0}$.) 
Let now $x_\alpha=\ln z_\alpha$ be a branch
of logarithm taking positive real values.  Let $Z_\alpha$
be the value of $x_\alpha$ at the point the disk $u_2$ leans on 
( other than the points $z_\alpha=0$ and $z_\alpha=\infty$)(cf. fig. 3 A,B).
\begin{equation}
{\epsfxsize16\baselineskip\epsfbox{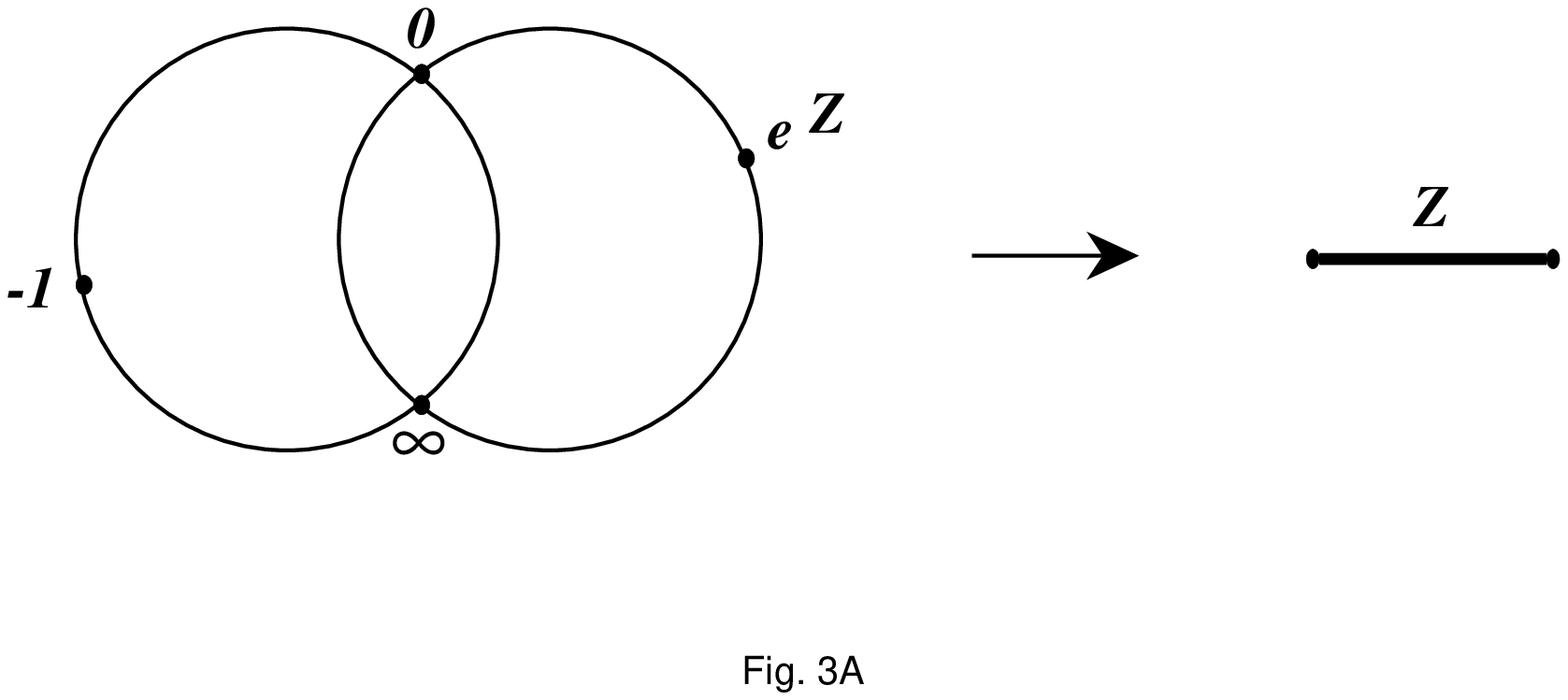}}
\nonumber
\end{equation}
\begin{equation}
{\epsfxsize16\baselineskip\epsfbox{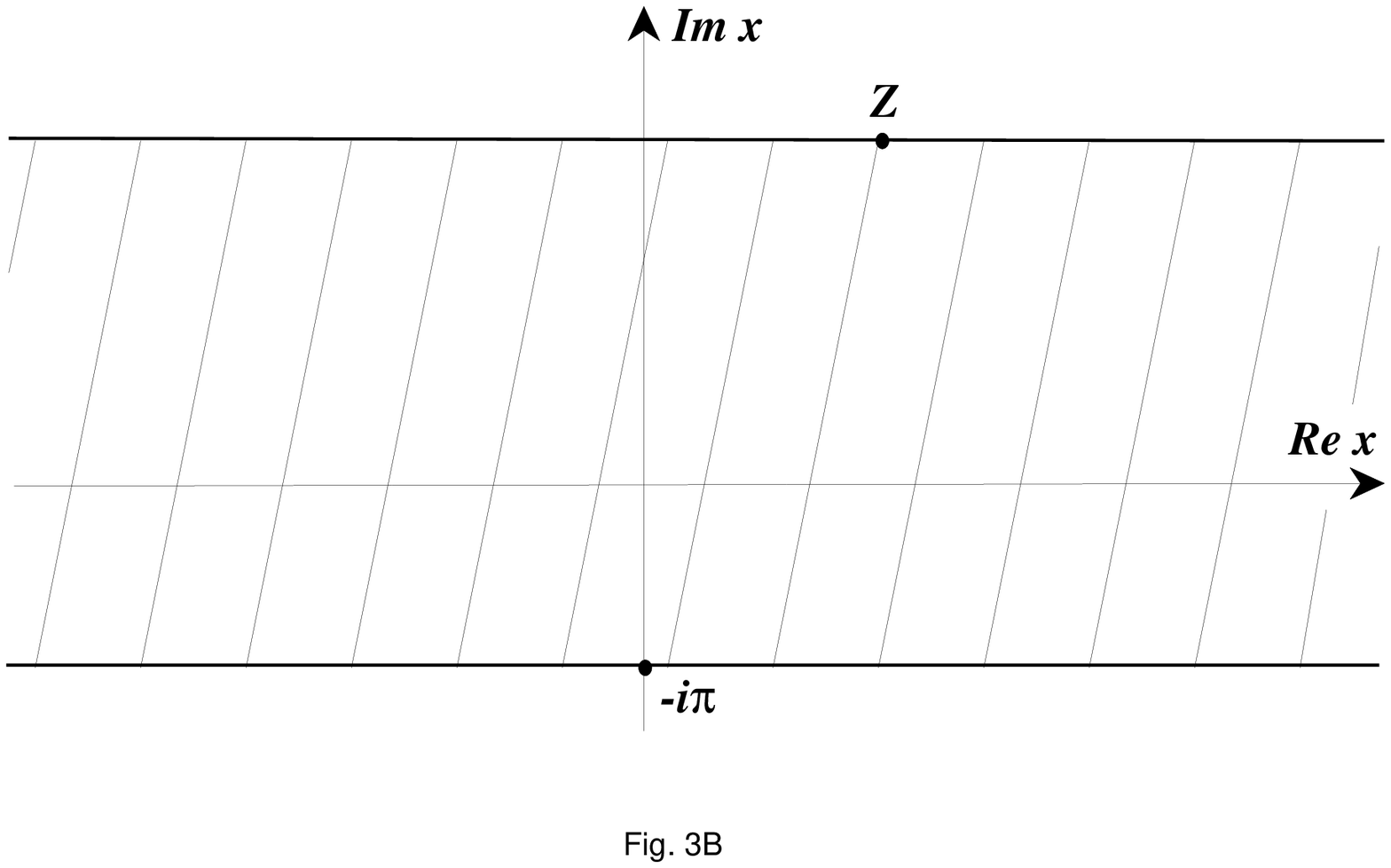}}
\nonumber
\end{equation}
This point is evidently outside
the disk $u_1$
and therefore $\mbox{Im} Z_\alpha > 0$. Assign this 
complex number $Z_\alpha$ with positive imaginary part to the edge $\alpha$. 
The complex number $Z_\alpha$ unambiguously determines a configuration 
of two disks leaning on four punctures. One can straightforwardly 
check that the definition of the number $Z_\alpha$ is correct:
\begin{proposition}
The value of $Z_\alpha$ is independent on the orientation of the edge
$\alpha$. The coordinate $x_\alpha$ changes to 
\begin{equation}
x_\alpha \mapsto Z_\alpha-i\pi-x_\alpha \label{Ori}
\end{equation}
under change of 
orientation of $\alpha$.
\end{proposition}

\subsection{A surface with projective structure from a fat graph}
\label{a}
Consider now the an inverse procedure i.e. how to restore a
surface starting from a graph with complex numbers assigned to the edges. 
It turns out that all the procedure from the above section can be performed
in the reversed direction.  Moreover it turns out that one is able to build a 
surface out of a graph with weaker condition on the numbers on edges, 
than that of positivity of the imaginary part. 
\begin{proposition} \label{constr}
For any fat graph with positive imaginary part complex numbers assigned
to the edges there exists a surface such that the above described construction
reproduces this graph.
\end{proposition}
{\em Proof} Assign a strip $-i\pi < \mbox{Im}x_\alpha < \mbox{Im}Z_\alpha$
to each oriented edge of the graph and define transition functions between 
the strips:
\begin{equation}
x_\alpha = Z_\alpha + \ln(e^{x_\beta} + 1) \label{Trans}
\end{equation}
\begin{equation}
x_{\alpha^\vee} = Z_\alpha - i\pi - x_\alpha
\end{equation}
where  $\alpha^\vee$ is an edge obtained from 
$\alpha$ by orientation changing. 
Here the orientations of edges are chosen in such a way that the end $v$
of $\alpha$ coincides with the beginning of $\beta$ and the edge $\alpha$
is next after the edge $\beta$ in counterclockwise direction w.r.t. $v$.
(fig 4). 
\begin{equation}
{\epsfxsize16\baselineskip\epsfbox{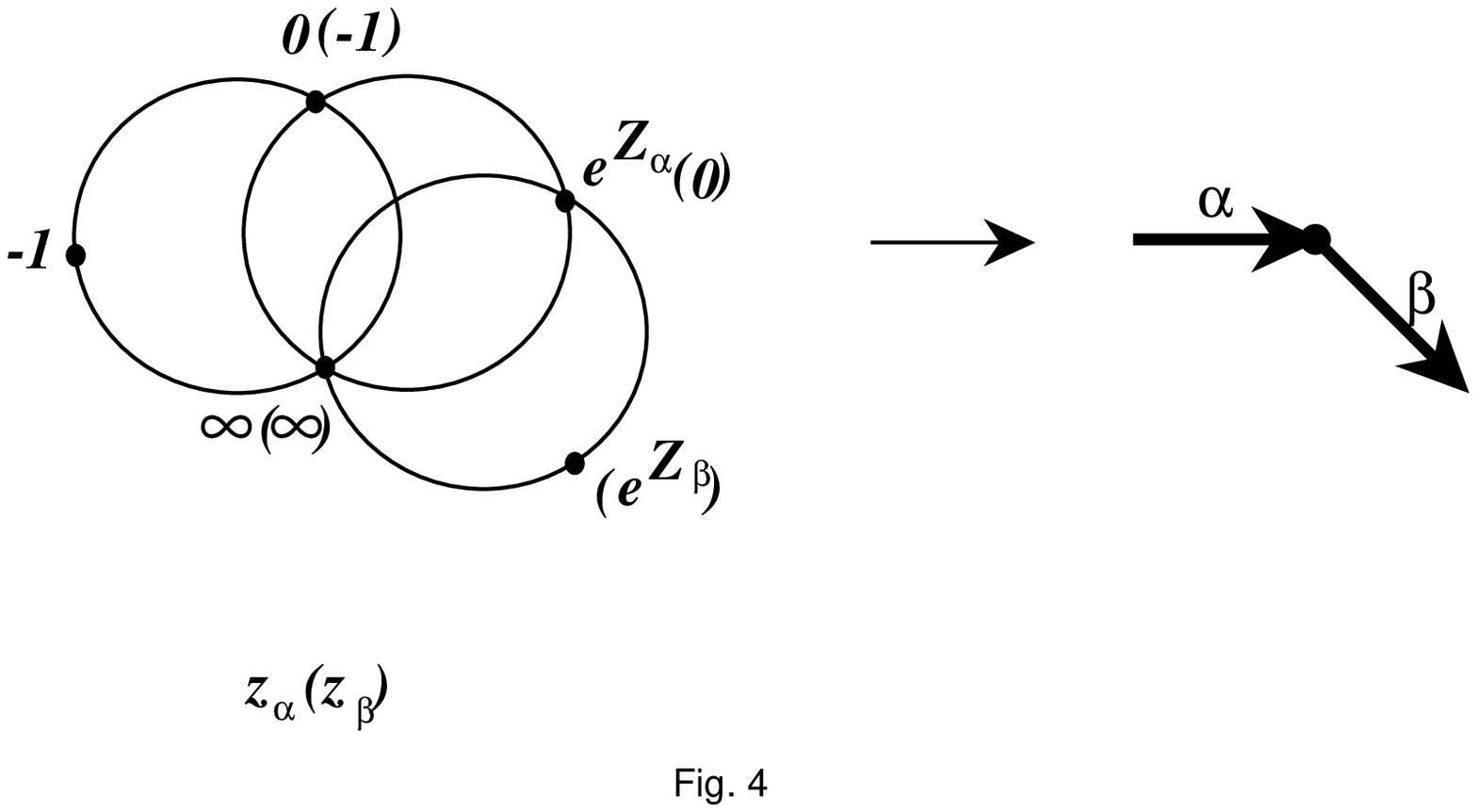}}
\nonumber
\end{equation}
The described set of strips and transition functions between them
defines the required surface provided that it is smooth. (For arbitrary set
of patches and transition functions between them the corresponding space
may be neither smooth manifold nor even Hausdorf topological space.) 
Remind, that formally the surface is defined by the set of strips and 
transition 
functions as a set of equivalence classes of the points of disjoint
union of strips; two points $a$ and $b$ are taken to be equivalent if 
there exist a chain of points $a=z_0,\ldots,z_n=b$ and of transition
functions $f_1,\ldots,f_n$ such that $z_{i+1} = f_i(z_i)$. The surface defined
by such rules is evidently smooth if (but not only if) each equivalence class
consists of finite number of points. We now prove, that this condition
holds in our case.

\noindent {\bf Lemma. } {\em Let $\alpha,\beta$ be two subsequent edges of 
the graph (subsequent means that the beginning of $\beta$ coincides with 
the end of $\alpha$. Then}
\begin{equation}
\mbox{Im} x_\alpha > \mbox{Im} Z_\alpha + \mbox{Im} x_\beta  \label{ineq}
\end{equation}

{\em Proof of the lemma:} For $\alpha$ next after $\beta$ in counterclockwise 
direction it follows directly from (\ref{Trans}) and from the fact that the
transition function is defined only for $-i\pi < x_\alpha < 0$.
 For $\beta$ next to
$\alpha$ combining (\ref{Trans}) and (\ref{Ori}) we get the analog of
(\ref{Trans}) for this case:
\begin{equation}
x_\alpha = Z_\alpha - \ln(e^{-x_\beta} + 1)
\end{equation}
One can easily check, that (\ref{ineq}) holds for this case also.
$\Box$

Now consider a sequence of equivalent points. Without loss of generality
it is sufficient to consider only such sequences for which 
$f_i \neq f_{i+1}^{-1}$
for all $i$ and let $\alpha_1,\ldots,\alpha_n$ be the corresponding
sequence of edges forming a path on the graph and oriented along this
path. Applying (\ref{ineq}) to the pairs of subsequent edges of this path
we get:
\begin{equation}
\mbox{Im} z_0 \geq \sum_{i=1}^{n}\mbox{Im}Z_{\alpha_i} + \mbox{Im} z_n
\end{equation}
which can be valid only for a finite sequence of edges. $\Box$

Note, that the proof of this theorem shows that a surface can be defined
starting from a graph with weaker condition on the numbers assigned to edges:
\begin{equation} \label{cond}
\sum_{\alpha \in \gamma} \mbox{Im} z_\alpha \geq 0
\end{equation}
Where $\gamma$ is any closed path on the graph without returns. ({\em Without
returns} means that it does not go along an edge forth and immediately back).  

Two different graphs with numbers satisfying (\ref{cond}) can correspond to 
the same projective structure (which never holds for graphs with positive 
imaginary part numbers).  In particular holds the following

\begin{proposition} \label{flip}
The graphs with numbers on the edges connected by the operation of 
flipping an edge (fig. 5) correspond to the same surface.
\end{proposition}

\begin{equation}
{\epsfxsize16\baselineskip\epsfbox{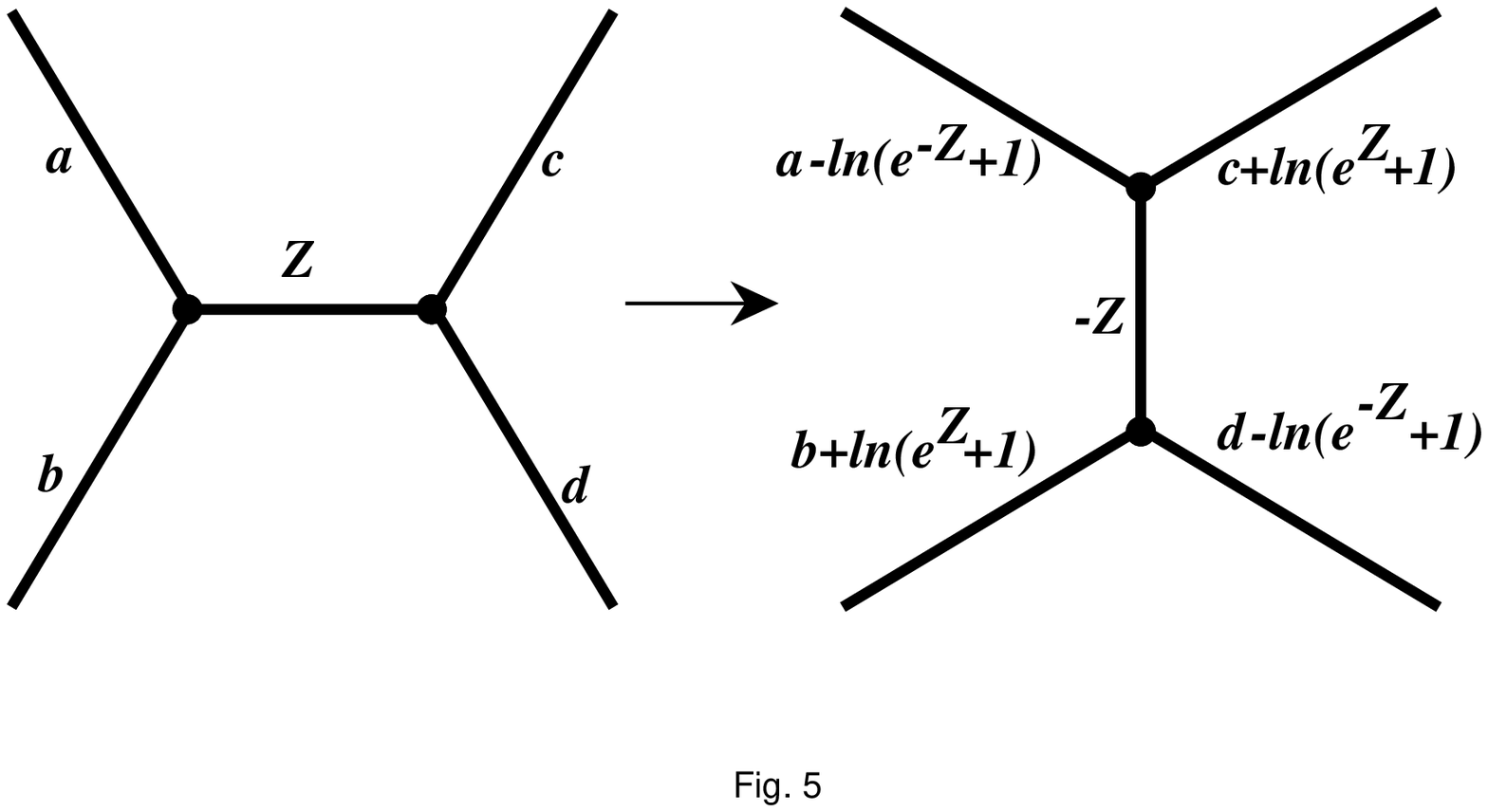}}
\nonumber
\end{equation}

The condition (\ref{cond}) allows us to extend the functions $Z_\alpha$
from the domain of  ${\cal MP}_\Sigma$ described by a given graph and the 
proposition (\ref{flip}) gives us the transition functions between 
the coordinates on ${\cal MP}_\Sigma$ which correspond to different graphs.

\section{Projective structures and flat $PGL(2,{\bf C})$ connections.}
\label{psafs}
Projective structures on a Riemann surface $\Sigma$ are closely related to
flat connections on a $PGL(2,{\bf C})$-bundle on $\Sigma$ (cf. \cite{BFK},
where these relations was discussed in terms of smooth $SL(2,{\bf C})$ flat
connections). Here we define a mapping 
\begin{equation} \label{conn}
{\cal MP}_\Sigma  \rightarrow {\cal A}_{\Sigma}(PGL(2,{\bf C}))
\end{equation}
where ${\cal A}_{\Sigma}(PGL(2,{\bf C}))$ is the moduli space of
flat structures of a $PGL(2,{\bf C})$-bundle or in other words the space
of flat $sl(2,{\bf C})$-connections on $\Sigma$ modulo gauge transformations.
Then we describe this mapping in terms of the coordinates $\{Z_\alpha\}$ on 
${\cal MP}_\Sigma$ and graph connections and also give in these terms 
a construction for the inverse (multivalued) mapping.

\subsection{Flat $PGL(2,{\bf C})$ connections from projective structures.}
\label{e}
Let $\Sigma$ be a surface equipped with a projective structure and 
$\{z_\alpha\}$ be a full set of projective coordinates on it. For each 
two of these coordinates $z_\alpha$ and $z_\beta$ with nontrivial intersection
of their definition domains one can define an element of the group 
$PGL(2,{\bf C})$($= GL(2,{\bf C})/{\bf C}^{*} = SL(2,{\bf C})/\{\pm 1\} =
(\mbox{group of M\"obius transformations.}$)) represented by a matrix 
(\ref{Mob})
\begin{equation}
g_{\alpha,\beta}= \left( \begin{array}{cc}
                                    a_{\alpha,\beta} & b_{\alpha,\beta} \\
                                    c_{\alpha,\beta} & d_{\alpha,\beta}
                                     \end{array}     \right)                 
\end{equation}
(Strictly speaking this matrix is defined unambiguously only for topologically 
trivial system of coordinate patches $\{z_\alpha\}$.
In general the matrix is defined if we have fixed also the connected component 
of the intersection of definition domains of $z_\alpha$ and 
$z_\beta$.) It is evident that this system of matrices satisfies the cocycle 
condition i.e. that for any three coordinates $z_\alpha,z_\beta,z_\gamma$
with intersecting definition domains one has
\begin{equation}
g_{\alpha,\beta}\/g_{\beta,\gamma}\/g_{\gamma,\alpha} = 1
 \in PGL(2,{\bf C})
\end{equation}
and thus this system of elements of $PGL(2,{\bf C})$ can be taken as a set 
of transition functions of a $PGL(2,{\bf C})$-bundle with 
canonical locally flat
connection.  (Flat sections w.r.t. this connection are those given by 
constant functions in the trivialization fixed by the transition functions.)  

The isomorphism class of a flat connection (the gauge 
equivalence class of a flat connection)
is fixed if one know the monodromy operators along a 
sufficient number of paths.
For example one can describe such class by fixing monodromy operators along 
edges of a graph drawn on the surface and homotopicaly equivalent to it
with, may be, some additional holes. The assignment of group elements
to oriented edges of a graph (in such a way that if one changes the orientation
of an edge the corresponding group element changes to its inverse) is called
{\em graph connection}.  Now we describe a  $PGL(2,{\bf C})$ graph connection
which corresponds to a given projective structure i.e. 
describe a procedure which
makes a graph connection on a graph starting from a 
projective structure, defined
by some (may be another) graph $\Gamma$ with complex numbers on edges. 
For our purpose it is convenient  to define the 
required graph connection on a graph
obtained from $\Gamma$ by blowing up vertices: The blown up graph 
$\tilde{\Gamma}$ is the graph $\Gamma$ with $k$-valent vertices replaced by 
$k$-vertex polygons (fig. 6A). 
\begin{equation}
{\epsfxsize16\baselineskip\epsfbox{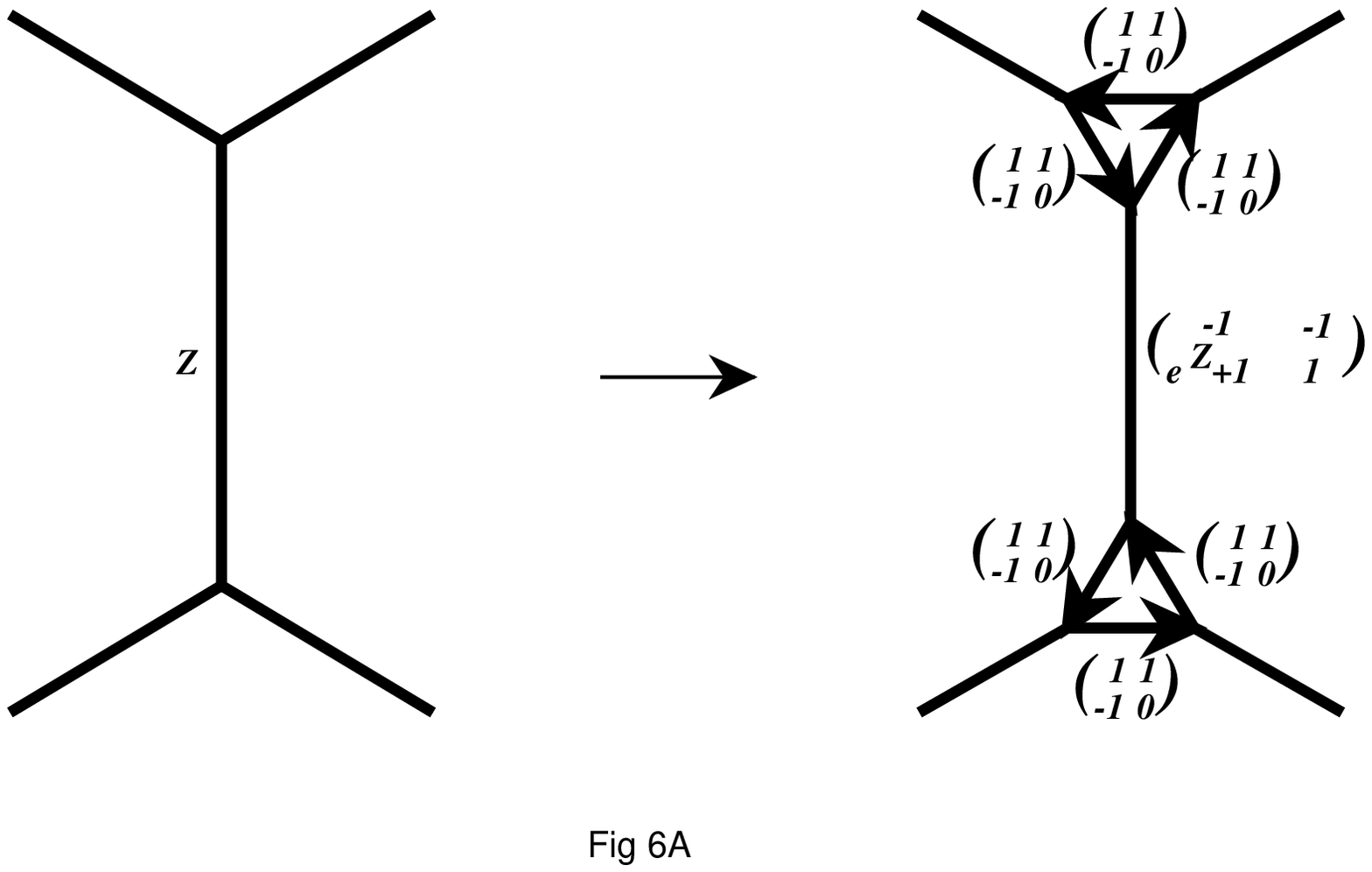}}
\nonumber
\end{equation}
Orient the new edges in counterclockwise
direction w.r.t the interior of the polygons and assign the matrices 
$g_\alpha =\left( \begin{array}{cc} 1 & 1 \\ -1 & 0 \end{array} \right) 
\in PGL(2,{\bf C})$
to them and  
$\left( \begin{array}{cc} -1 & -1 \\  e^{Z_\alpha} + 1 & 1 \end{array} \right)
 \in PGL(2,{\bf C})$ to the old edges $\{ \alpha \}$ (their orientations 
are inessential because $g_\alpha = g_\alpha^{-1}$). 
\begin{proposition}
The above described graph connection (fig. 6A) on $\tilde{\Gamma}$ corresponds 
to the projective structure defined by the graph $\Gamma$ and the numbers
$\{ {Z_\alpha} \}$.
 \end{proposition}

For practical purposes it is often more convenient to 
describe flat connections 
in terms of another graph $\tilde{\tilde{\Gamma}}$ --- the graph $\Gamma$ 
with blown up edges and vertices. This graph can be obtained from the graph 
$\Gamma$ by replacing its edges by rectangulars (fig. 6B). 
\begin{equation}
{\epsfxsize16\baselineskip\epsfbox{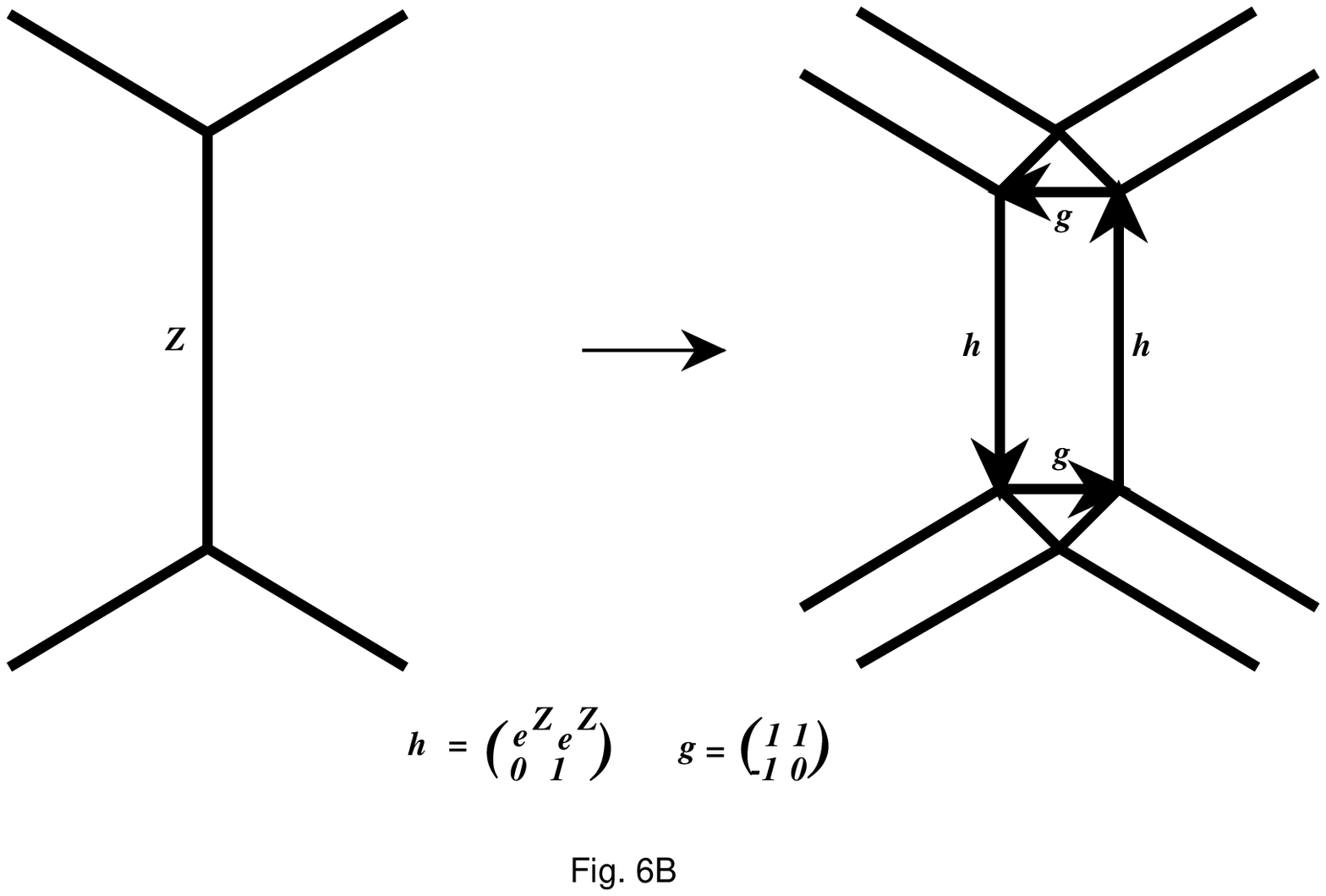}}
\nonumber
\end{equation}
Let us call the 
 {\em long} edges the edges parallel to the edges of the original 
graph $\Gamma$ and {\em short} edges the other ones. Orient the 
edges in counterclockwise direction w.r.t. the interior of the 
rectangulars and assign the matrices 
$\left( \begin{array}{cc} e^{Z_\alpha} & e^{Z_\alpha} \\ 0 & 1 
\end{array} \right)$ to the long edges and 
$\left( \begin{array}{cc} 1 & 1 \\ -1 & 0 \end{array} \right) $ to the short 
ones. Then the obtained graph connection on $\tilde{\tilde{\Gamma}}$ 
defines an element of  ${\cal A}_{\Sigma}(PGL(2,{\bf C}))$ which 
corresponds to the graph $\Gamma$ and the set of numbers $\{Z_\alpha\}$ 
on edges.

\subsection{Projective structures from flat $PGL(2,{\bf C})$ connections.}
\label{d}
Now consider the problem, what is the preimage of a flat connection.
In order to solve it we give an explicit construction of all graphs with
numbers corresponding to a given graph and to a given 
$PGL(2,{\bf C})$-graph connection on it.

Let $\Gamma$ be a threevalent graph homotopically equivalent to $\Sigma$ 
and $g_{\alpha}$ be a graph connection (i.e. group elements
assigned to the edges). Consider a monodromy operator  $g_{v,f}$ around a 
face $f$ starting and ending at a vertex $v$ (i.e. a path ordered product
of group elements assigned to the sites of the face $f$.   Choose an 
eigenspace $l_{v,f} \subset {\bf C}^2$ for each $g_{v,f}$ in such a way,
that the monodromy operator around $f$ sends $l_{v_1,f}$ to $l_{v_2,f}$ 
for any  corners $v_1$ and $v_2$ of $f$ (i.e. one has two possible choices
for each face). Thus for each vertex we have three eigenspaces --- one for
each face this vertex is a corner of. Now consider an edge $\alpha$ of the 
graph. The monodromy operator along the edge sends two of the three 
eigenspaces assigned to one end of the edge onto two eigenspaces assigned 
to another one. The image of the third eigenspace together with the 
eigenspaces assigned to this end form a quadruple of one dimensional 
subspaces in ${\bf C}^2$ (or they can be thought of as a quadruple of points 
in ${\bf CP}^1$ which we denote as $P_{-1}, P_0, P_\infty$ and $P$). 
Such quadruple is known to have one $GL(2,{\bf C})$-invariant --- the 
double ratio: 
\begin{equation}
z=-\frac{(P_0 - P)}{(P_\infty - P)} 
        \frac{(P_\infty - P_{-1})}{(P_0 - P_{-1})}
\end{equation}
Let $Z_\alpha$ be any positive imaginary part value of $\ln z$.  Thus we 
have assigned a positive imaginary part complex number to all edges of the
graph.
\begin{proposition}
The $PGL(2,{\bf C})$-flat structure which corresponds to the projective 
structure determined by the constructed graph with complex numbers
coincides with that we have started from and all sets of numbers assigned
to edges can be obtained in this way.
\end{proposition}
The proof can be given by a direct verification.

This construction shows that  the inverse image of a 
$PGL(2,{\bf C})$-flat structure is a set the elements of which are numerated 
by different ways to choosing the graph, the eigenspaces an the branches 
of logarithm. 
At a generic point where the monodromy operators around all faces are 
diagonalizable and not equal to unity this set is discrete and thus the 
mapping (\ref{conn}) is a covering (with infinite number of sheets).
For the points, 
where the monodromy operator around at least one face is not diagonalizable 
this covering is not locally trivial. For the points where 
there exist faces the 
monodromy operators around which equal to unity the inverse image is not 
discrete.

Therefore we have described the mappings of the diagram
\begin{equation}
\setlength{\unitlength}{0.3em}
\begin{picture}(17,28)(10,-5)
\put(20,20){\makebox(0,0){${\cal MP}_\Sigma$}}
\put(0,0){\makebox(0,0){${\cal MP}_\Sigma^{comb}$}}
\put(42,0){\makebox(0,0){${\cal A}_{\Sigma}(PGL(2,{\bf C}))$}}
\put(1,4){\vector(1,1){13}}
\put(15,16){\vector(-1,-1){13}}
\put(23,17){\vector(1,-1){13}}
\put(7.5,0){\vector(1,0){21}}
\put(28.5,1){\vector(-1,0){21}}
\put(9,7){$\ref{a}$}
\put(4,12){$\ref{b}$}
\put(31,12){$\ref{psafs}$}
\put(16,1.5){$\ref{d}$}
\put(16,-3){$\ref{e}$}
\end{picture}
\end{equation}
and proved its commutativity. (The numbers indicate section  
where the corresponding mapping was considered.)

\section{Examples and unsolved problems.}

\subsection{Poincar\'e projective structure.}
 The construction  from sect. \ref{a} which makes a graph starting from the
projective structure is not applicable for this case because
the maximal disks w.r.t. the Poincar\'e projective structure pulled back on
the universal covering of a curve are just the mappings onto the universal 
covering and thus all maximal disks lean on all punctures 
infinitely many times.
Nevertheless the inverse construction may give Poincar\'e projective structure:

\begin{proposition}
Poincar\'e projective structure corresponds to graphs with real numbers
assigned to edges (via construction of sect \ref{a}). 
\end{proposition}

{\em Proof.} One can easily see that the coordinate patches for the case of
real numbers assigned to edges are unit disks (in terms of coordinates
$z_\alpha$) and the transition functions maps one disc onto another. 
In particular it means that the corresponding $PGL(2,{\bf C})$-bundle 
reduces to $PGL(2,{\bf R})$ one and that the procedure of gluing  strips 
reduces to factorization of a single disk thus giving just the Poincar\'e 
uniformization mapping. Since the surface which can be constructed starting
from a graph with positive imaginary parts numbers assigned to edges is
smooth the same is true for graphs with real numbers on edges. $\Box$

Note, that the correspondence between surfaces with Poincar\'e projective
structure and fat graphs with real numbers on edges is not one-to-one.
In particular the graphs connected with each other as shown on 
fig. 5 correspond
to the same surface. I seem to be probable, that all graphs corresponding  the
same surface  can be obtained one from another by such operation.

Nevertheless  this construction gives us at least local parameterization of 
the Poincar\'e projective structures and thus of the moduli space of complex 
structures. In these terms it is easy to find a Fuchsian group, corresponding
to a given complex structure.

\begin{proposition}
The complex surface corresponding to a given graph with real numbers
on the edges is isomorphic to the quotient of a unit disc by the monodromy
group of the flat connection shown on fig. 6 A,B.
\end{proposition}

This proposition follows immediately from prop. \ref{constr}. Note, that
{\em a priori} it was not evident, that the Fuchsian group, given by this
construction always corresponds to a smooth surface.
 
\subsection{Covering projective structure.}
Covering projective structures are in a sense the most simple ones, for
which it is possible to construct the corresponding graphs in a very simple 
explicit way.  The inverse operation -- to restore
the ramification points and the scheme of the covering starting from a graph --
can also be done explicitly, provided the graph indeed corresponds
to a covering projective structure. Here we give a simple proposition which
allows to characterize such graphs.
\begin{proposition}
Covering  projective structure is characterized by the requirement  that the
corresponding  $PGL(2,{\bf C})$-connection is trivial.
\end{proposition}
 One can easily write 
down this condition in terms of the variables $Z_\alpha$ using the construction
of sect. \ref{e}. In particular a projective structure at a neighborhood of a 
puncture is isomorphic to that at a neighborhood of a $k$-th order ramification
point iff
\begin{equation} \label{cov}
\left\{ \begin{array}{l} Z_{\alpha_1}+\ldots+Z_{\alpha_n} = 2\pi i k \\
                              e^{Z_{\alpha_1}}+e^{Z_{\alpha_1}+ Z_{\alpha_2}}+
                              \ldots +e^{Z_{\alpha_1}+\ldots+Z_{\alpha_n}} = 0 
              \end{array} \right.
\end{equation}
where $\{\alpha_i\}$ is the sequence of sites of the face, corresponding 
to the given puncture. In particular $k=1$ means that the projective structure
at the puncture is nonsingular. For graphs which correspond to punctured
spheres the equations (\ref{cov}) satisfied for all faces of the 
graph is enough
for the corresponding projective structure to be a covering one. For surfaces 
with handles we have to impose additional conditions, which can be written
down explicitly for each concrete graph.

\subsection{Relations to Strebel construction}
In this section we discuss the relation between our construction and that of 
Penner/Stre\-bel which describes a one-to-one correspondence between 
the space of {\em complex} structures on a surface $\Sigma$ and the 
space of graphs with positive real numbers assigned to edges.

Let $\Gamma$ be a graph with positive real numbers $l_\alpha$ assigned 
to edges and let  $\tilde{\Gamma}$ be the corresponding graph with blown 
up vertices.  Assign the purely imaginary complex numbers $i(l_\alpha - \pi)$ 
to the edges which correspond to the edges $\alpha$ of the old graph and 
$i\pi$ to all other edges of $\tilde{\Gamma}$ (fig. 10).
\begin{equation}
{\epsfxsize16\baselineskip\epsfbox{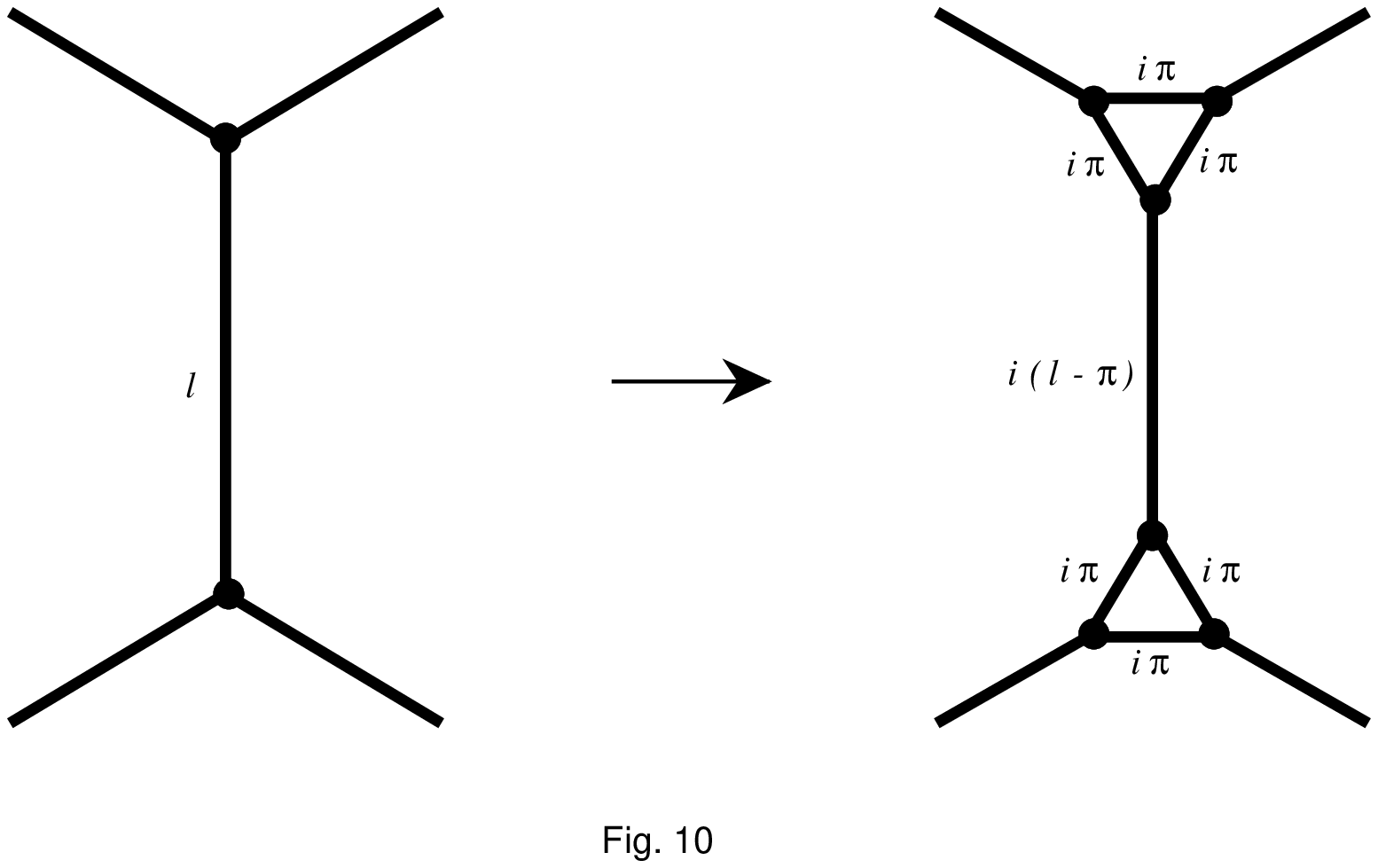}}
\nonumber
\end{equation}
 A surface with projective 
structure which corresponds to this graph with numbers by the construction 
of sect. \ref{b} (it is applicable here because the condition (\ref{cond}) is 
satisfied) has two kind of punctures: the punctures of the first kind 
correspond to the faces of $\Gamma$ and that of the second one --- to the 
vertices of $\Gamma$. Extend the complex structure to the punctures 
of the second kind. 
\begin{proposition} 
The complex surface obtained by such construction starting from a graph
with positive real numbers on edges gives the complex surface isomorphic to 
that given by Strebel/Penner construction.
\end{proposition}

The proof of the proposition can be done by direct comparison of our 
construction of a surface (sect \ref{a}) and the Strebel one \cite{St}.

\subsection{Projective structures on the torus with one puncture.} \label{pstp}
Consider as  an example the moduli space of projective structures on a torus 
with one puncture and with nonsingular behavior of the projective structure at
this puncture. Such projective structure can be characterized by two 
parameters: the standard modular parameter $\tau$ of complex structure
and a parameter $k$ for projective connection $T = k^2dz^2$, where $z$ is
is the standard coordinate on the torus ${\bf C}/{\bf Z}^2$ 
$(z \equiv z + m + n\tau)$.
The projective coordinates are given therefore by ratios of solutions of
the equation (\ref{Hil}), i.e a general projective coordinate $u$ has the form
\begin{equation}
u = \frac{ae^{kz} + be^{-kz}}{ce^{kz} + de^{-kz}}~,~\mbox{ for } k \neq 0.
\end{equation}
\begin{equation}
u = \frac{az + b}{cz + d}~,~\mbox{ for } k = 0.
\end{equation}
For $k=0$ the corresponding maximal disks are shown on fig. 9.
\begin{equation}
{\epsfxsize16\baselineskip\epsfbox{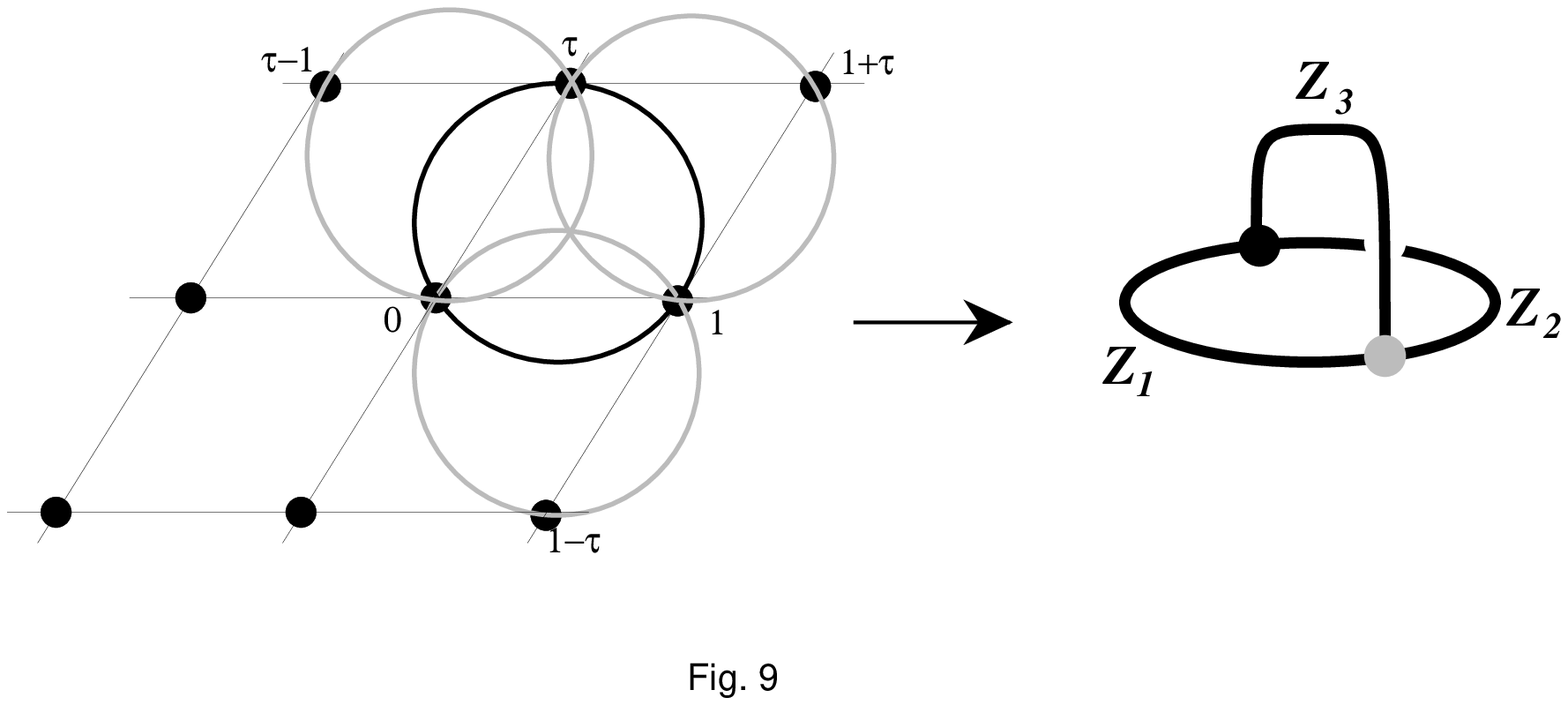}} \nonumber
\end{equation}
 There are two
inequivalent disks corresponding to two vertices of 
the graph and three ways to 
deform one into another, corresponding to the three edges of the graph. For 
$k \neq 0$ the the disks are deformed (on the $z$-plane), but topologically
the picture remains unchanged. To calculate the numbers on edges
one has to take logarithms of double ratios of values of any projective 
coordinates at quadruples of punctures:
\begin{eqnarray}
Z_1 = \ln \left(-\frac{u(\tau)-u(1+\tau)}{u(1) - u(1+\tau)}\right)
                    \left(\frac{u(1)-u(0)}{u(\tau)-u(0)}\right) =
                     2\ln \left(-\frac{\sh k}{\sh k\tau}\right)\\
Z_2 = \ln  \left(-\frac{u(1)-u(1-\tau)}{u(0) - u(1-\tau)}\right)
                    \left(\frac{u(0)-u(\tau)}{u(1)-u(\tau)}\right) =
                     2\ln \left(-\frac{\sh k\tau}{\sh k(\tau-1)}\right)\\
Z_3 = \ln  \left(-\frac{u(0)-u(\tau-1)}{u(\tau) - u(\tau-1)}\right)
                    \left(\frac{u(\tau)-u(1)}{u(0)-u(1)}\right) =
                     2\ln \left(-\frac{\sh k(\tau-1)}{\sh k}\right)
\end{eqnarray}

\subsection{Projective structure on sphere with four nonsingular punctures.}
The construction of a graph for a standard projective structure on a sphere 
with four punctures is illustrated in fig. 8.
\begin{equation}
{\epsfxsize16\baselineskip\epsfbox{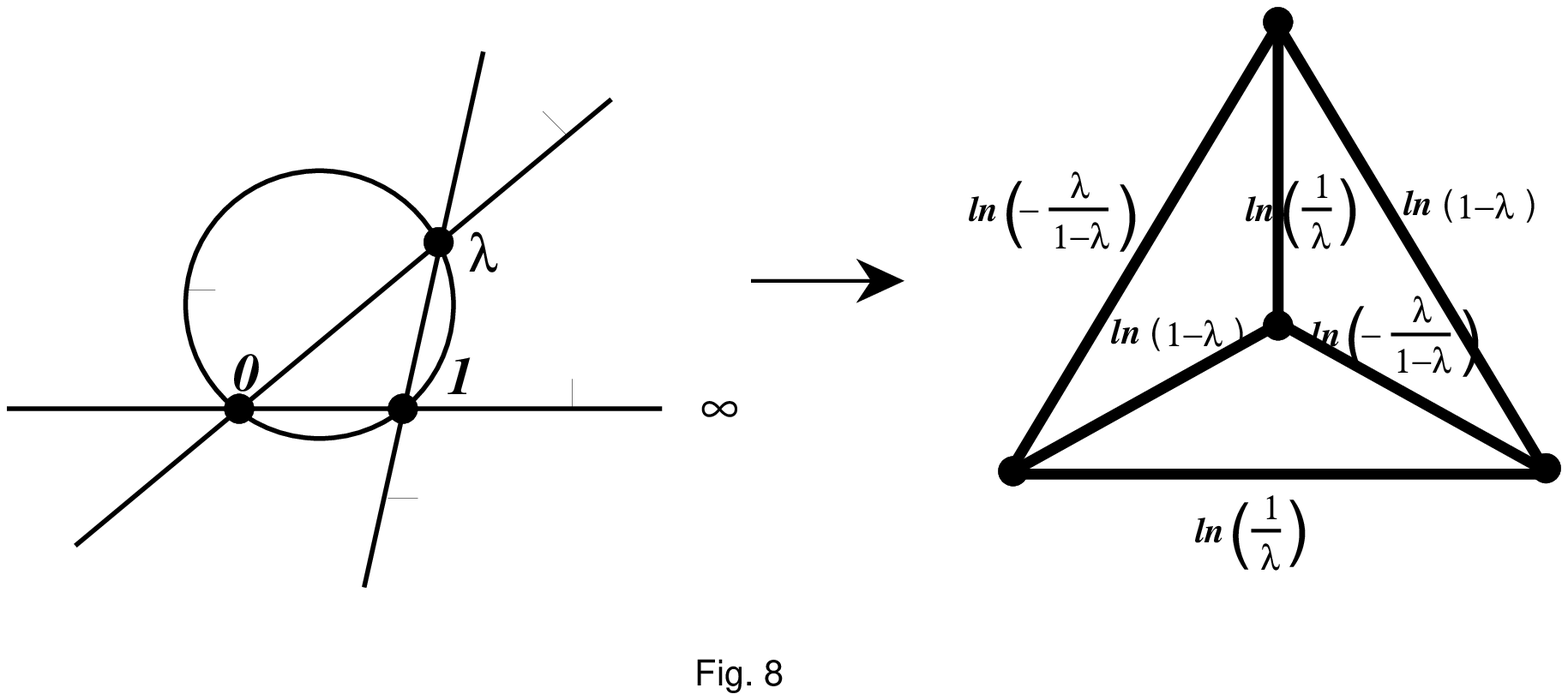}}
\nonumber
\end{equation}
 The procedure for this case as well 
as for all other coverings of the Riemann sphere reduces to calculating some 
double ratios of coordinates of the ramification 
points and nonsingular punctures. 

\subsection{Unsolved problems.}
To conclude we discuss some yet unclear aspects of the above construction. 
First of all, we do not have a rigorous proof of  
the fact that our construction really 
describes the dense subset of  ${\cal MP}_\Sigma$, though this conjecture 
seems to be very realistic. Another yet unclear related point is how to 
describe the transition functions between the maps of ${\cal MP}_\Sigma$, 
corresponding to different graphs. The proposition \ref{flip} shows how to do 
this in some cases, but for example, the question, what is glued to the 
component of the boundary of a cell corresponding to a graph with 
closed edge and real numbers on this edge 
remains unclear. The example \ref{pstp} shows that when the projective 
structure tends to the Shottky one ($k \rightarrow i\pi$ in our example) some 
numbers $Z_\alpha$ tend to infinity. The question is, how to generalize the 
space ${\cal MP}_\Sigma^{comb}$ to be able to describe 
all projective structures.

Another group of problems are connected to the perspective of possible 
quantization of the space ${\cal MP}_\Sigma$. The space ${\cal MP}_\Sigma$ is
a Poisson 
manifold and the first question is, how to express the 
Poisson structure in terms 
of the coordinates $Z_\alpha$.\footnote{Recently this 
problem was solved by O.Zaboronsky and the author and will 
appear in a forthcoming publication.} 
Another problem is how to describe the set of algebraic 
functions on ${\cal MP}_\Sigma$. The knowledge of such 
description is necessary 
for applying some algebraic quantization technique, 
like quantum groups e.t.c., 
analogously to what can be done for the space  
${\cal A}_{\Sigma}$ (see \cite{FR}). 

\begin{thebibliography}{9}
\bibitem{P}\ R. C. Penner, {\em The decorated Teichm\"uller space of 
punctured surfaces,} Commun. Math. Phys., {\bf 113}(1987),299
\bibitem{St}\ K.Strebel, {\em Quadratic Differentials}, Springer--Verlag, 1984.
\bibitem{K}
M.L.Kontsevich, {\sl Funk.Anal.\&Prilozh.}, {\bf 25} (1991) 50 (in Russian);
{\it Intersection Theory on the Moduli Space of Curves and the Matrix
Airy Function}, {\sl Comm.Math.Phys.}, {\bf 147} (1992) 1.
\bibitem{Kon91}\ M.Kontsevich {\it Formal non-commutative Symplectic
 geometry.}M.P.I. Preprint (1992).
\bibitem{F}\ Whittaker, E.T., Watson, G.N.{\em Course of modern
analysis}, 4th edition, Cambridge Univ. Press, 1958
\bibitem{Ch}\ L Chekhov, {\em Matrix Models: a Way to Quantum 
Moduli Spaces}, preprint PAR--LPTHE--93--22.
\bibitem{BFK}\ A.Bilal, V.V.Fock and I.I.Kogan, {\em On the origin of
 $W$-algebras.}, Nucl.Phys. {\bf B359}(1991)2-3,635-672.
\bibitem{FR}\ V.V.Fock and A.A.Rosly, {\em Poisson structure on
moduli space of flat connection and classical $r$-matrix}, 
preprint ITEP-72-92
\end{thebibliography}
\end{document}